\documentclass[12pt]{iopart}
\usepackage{iopams}  
\usepackage[separate-uncertainty=true]{siunitx}
\DeclareSIUnit\gauss{G}
\usepackage[caption=false]{subfig}
\usepackage{multirow,float,graphicx}
\usepackage{bm,mathrsfs,accents}
\usepackage{pdflscape}

\begin{document}

\title[Review and experimental benchmarking of machine learning optimization]{Review and experimental benchmarking of machine learning algorithms for efficient optimization of cold atom experiments}

\author{Oliver Anton$^1$\footnotemark{}, Victoria A. Henderson$^{1}$\footnotemark[\value{footnote}], Elisa Da Ros$^{1}$, Ivan Sekulic$^{2,3}$, Sven Burger$^{2,3}$, Philipp-Immanuel Schneider$^{2,3}$, Markus Krutzik$^{1,4}$\footnotetext{O.A. and V.A.H. are joint first authors.}}

\address{$^1$ Institut f\"{u}r Physik and IRIS, Humboldt-Universit\"{a}t zu Berlin, Newtonstr. 15, Berlin 12489, Germany}
\address{$^2$ JCMwave GmbH, Bolivarallee 22, 14050 Berlin, Germany}
\address{$^3$ Zuse Institute Berlin (ZIB), Takustraße 7, 14195 Berlin, Germany}
\address{$^4$ Ferdinand-Braun-Institut (FBH), Gustav-Kirchoff-Str.4, 12489 Berlin}
\ead{oliver.anton@physik.hu-berlin.de}
\vspace{10pt}

\begin{abstract}
The generation of cold atom clouds is a complex process which involves the optimization of noisy data in high dimensional parameter spaces.
Optimization  can be challenging both in and especially outside of the lab due to lack of time, expertise, or access for lengthy manual optimization.
In recent years, it was demonstrated that machine learning offers a solution since it can optimize high dimensional problems quickly, without knowledge of the experiment itself.
In this paper we present results showing the benchmarking of nine different optimization techniques and implementations, alongside their ability to optimize a Rubidium (Rb) cold atom experiment.
The investigations are performed on a 3D $^{87}$Rb molasses with 10 and 18 adjustable parameters, respectively, where the atom number obtained by absorption imaging was chosen as the test problem. We further compare the best performing optimizers under different effective noise conditions by reducing the Signal-to-Noise ratio of the images via adapting the atomic vapor pressure in the 2D+ MOT and the detection laser frequency stability. 
\end{abstract}

%
%
%
%
%

\section{Introduction} \label{sec:Intro}

Cold atom systems are an essential part of the quantum revolution, facilitating new generations of quantum technologies. 
Although the theory behind laser and evaporative cooling techniques is well understood~\cite{Dalibard1989, metcalf1999, Ketterle1996, Ketterle1999}, in practice each apparatus is unique and optimal parameters must be found experimentally, typically in a lengthy manual optimization routine involving a large number of inter-dependent parameters.
Such a labour intensive approach requires a large amount of technical expertise and patience, and will often only find a local efficiency maximum due to the size and complexity of the parameter space.
Thus, by instead optimizing via machine learning, one can improve the productivity of cold atom experiments and also take steps towards autonomous operation or control by non-specialist operators.
This will be particularly essential for mobile quantum technologies such as gravimeters~\cite{Freier2016, Stray2022, Wu2019, menoret2018, antoni2022} and lattice clocks ~\cite{poli2014, Grotti2018, takamoto2020, gellesch2020, boulder2021} where the environmental conditions vary, or space missions on manned and unmanned spacecraft~\cite{elliott2018, Frye2019, devani2020, aveline2020, Sidhu2021, Ahlers2022, Alonso2022, DaRos2023}. 
However, the optimization of problems involving noisy data and large parameter sets is computationally heavy and requires the use of an appropriate optimization algorithm.

The topic of optimization has been of significant interest in the community, with a variety of approaches being taken.
Earlier examples utilised heuristic optimization algorithms, such as differential evolution algorithms, to maximize atom number in a magnetic trap loaded from a magneto-optical trap (MOT) ~\cite{Rohringer2008, Geisel2013} and demonstrated a higher atom number than that achieved by manual optimization.  
Here up to 21 parameters were optimized in \SI{5}{\hour}~\SI{45}{\minute} or \num{5000} evaluations.
Later, neural networks were used to optimize optical depth in a MOT by optimizing 63 parameters~\cite{Tranter2018}, and to optimize an evaporation ramp for $^{87}$Rb Bose-Einstein Condensate (BEC) using optical depth (OD) as indicator for the phase space density~\cite{Wu2020}.
Reinforcement learning is used in~\cite{reinschmidt2023} to optimize atom number and temperature simultaneously, as well as preparing a defined number of atoms in the MOT. The optimisation technique is performed on both a simulation and a real world experiment, using laser detuning as a single optimization parameter.
Currently the most popular machine learning (ML) technique within the field is Bayesian optimization (BO) using Gaussian process regression~\cite{Wigley2016, Nakamura2019, Barker2020, Davletov2020, Ma2023}.
Such routines were, for example, used to optimize the evaporation ramp for a crossed optical dipole trap (cODT) via a 16 parameter ramp. In this case, the minimization of the number of atoms in the wings of the cloud was used as an optimization goal~\cite{Wigley2016}. They were also able to identify which parameters are important to optimize and which could be excluded from the optimization as unimportant.
Another work benchmarks a different Bayesian optimizer against a random search, and shows the relative importance of different stages of an evaporation ramp for atoms in a cODT~\cite{Nakamura2019}; here 8 parameters are used to optimize the number of atoms in a circular area of interest, with 300 trials requiring~\SI{3}{\hour}, and repeating the optimization \num{16} times to reduce the effects of noise.
Further benchmarking work has occurred in~\cite{Barker2020}, comparing BO to neural networks and differential evolution by optimizing 37 parameters for cooling and subsequent evaporation to Bose-Einstein condensate (BEC) in a time-averaged orbiting potential trap;  this showed the significant advantage of BO over neural networks and differential evolution.  

While these results are promising in terms of applying BO for the preparation of ultra-cold atom systems, more systematic comparisons between different methods are required to facilitate the choice of the most suitable algorithms. 
BO as well as other ML methods can have a significant computational overhead for calculating the next parameter sample due to the required training and evaluation of the underlying ML model. 
On one hand, this calls for an efficient implementation of ML methods. 
But on the other hand, a quantification of the overhead versus the runtime of the cold-atom experiment is required to decide between ML methods and less computationally expensive heuristic methods. 

Another major challenge that influences the choice of the best optimization method is the inherently noisy data that is produced in the changing environments, especially for mobile experiments. 
Atom numbers and temperatures are typical optimization objectives. They are calculated from the fitting of the atom cloud absorption profiles. This fitting can be very sensitive to fluctuations, especially at low atom numbers.
This can degrade the performance of algorithms to varying degrees, depending on the underlying strategy. This difficulty can be tackled in a range of ways.
For example, one can take the average of several optimization cycles~\cite{Nakamura2019}.
An alternative approach is to adjust the algorithm, for example by resampling old population members in a differential evolution algorithm~\cite{Geisel2013}, or by using a cost function which is less sensitive to noise~\cite{Barker2020, Nakamura2019, Wigley2016}.
Many of the alternative cost functions have the additional benefit of requiring less computation time by using parameters such as optical depth as an indicator for atom number or phase space density.

In this paper, we study and benchmark how noise and dimensionality influences the performance of a large set of heuristic and ML optimization methods.
We consider the heuristic methods particle swarm optimization (PSO)~\cite{poli2007,kaveh2014} and its adaptation LILDE~\cite{Geisel2013}, differential evolution (DE), covariance matrix adaptation evolution strategy (CMA-ES)~\cite{CMAES} and downhill simplex, also known as Nelder-Mead search~\cite{nelder1965}.
As a baseline uninformed method, we include random sampling (RS) from a uniform distribution.
In addition, machine learning-based BO in various implementations~\cite{AX,FMFN} is considered, with particular emphasis on algorithms that can handle noise in an explicit manner. 
Moreover, an extension of BO is studied, which is explicitly designed to cope with noisy data without the need for repeated optimization or resampling~\cite{letham2019}. 
Although the extended BO method is more elaborate, we developed an efficient implementation that facilitates the optimization of a large number of parameters using a fitted atom number as a cost function with a small computational overhead~\cite{optimal}.

\section{Optimization algorithms for tuning experimental parameters}
\label{sec:optimizers}
There is a large number of algorithms that are suitable for minimizing different kinds of objective functions.
The laboratory experiments considered in this work can be regarded as a function $f: \bi{p} \in \mathcal{X} \subset \mathbb{R}^d \mapsto \mathbb{R}$ that maps a $d$-dimensional vector of experimental control parameters $\bi{p}$ to the objective value $f(\bi{p})$. The training data fed to the minimization methods $\tilde f(\bi{p}) = f(\bi{p}) + \epsilon$ is corrupted by a random noise $\epsilon$ that is assumed to have zero mean. In many cases, its probability distribution is well modelled by a normal distribution with a constant variance $\eta^2_{\rm noise}$, i.e. $\epsilon \sim \mathcal{N}(0,\eta^2_{\rm noise})$. The goal is to find the parameter vector $\bi{p}_{\rm min}$  that minimizes $\mathrm{E}[\tilde f(\bi{p})] = f(\bi{p})$ in the search space $\mathcal{X}$. The function $f(\bi{p})$ is in general non-convex and derivative information is unattainable, rendering gradient-based methods inappropriate. Since $d$ has a moderately large value (here $d=10$ or $d=18$) and the objective is expensive to evaluate, an exhaustive search in the full space $\mathcal{X}$ is also infeasible. In such a case, one often resorts to heuristic minimization strategies. These do not guarantee convergence to a global minimum, however they can explore larger parameter spaces and are to a certain degree robust towards noise.

\subsection{Heuristic minimization methods}
\label{sec:Heuristic}

All considered heuristic strategies start from a random population of $N$ vectors which is updated in each iteration. 

Particle swarm optimization (PSO) is a global minimization strategy that draws its inspiration from the field of sociobiology and the intelligence of animal groups (e.g. a flock of birds) \cite{zhang2015}. The swarm members move with individual velocities through the parameter space. These velocities are updated by a randomized weighted sum of the current velocities, velocities directed to the individual best seen parameters $\bi{p}_{{\rm best}, i}$, $i=1,\dots,N$, and the best seen parameters of the whole swarm $\bi{p}_{\rm best, swarm}$.

Differential evolution (DE) is an evolutionary optimization algorithm inspired by the mutation, crossover and selection processes occurring in nature~\cite{das2010}. In the mutation step, for each member $\bi{p}_i$ of the population, a mutated genome is created as $\bi{p}_{\rm mut} = \bi{a} + F(\bi{b} - \bi{c})$, where $F$ is the differential weight and $\bi{a}, \bi{b}, \bi{c}$ are distinct randomly selected population members. Random entries of $\bi{p}_i$, selected according to a crossover probability, are replaced with the mutated genome $\bi{p}_{\rm mut}$, forming a new candidate. The candidate replaces $\bi{p}_i$ in the next population if its objective value is lower.

The covariance matrix adaptation evolution strategy (CMA-ES) is another evolutionary algorithm~\cite{hansen2006}. It draws its $N$ population members from a multivariate normal distribution $\mathcal{N}(\bm{\mu}, \Sigma)$ with mean vector $\bm{\mu} \in \mathbb{R}^d$ and covariance matrix $\Sigma\in\mathbb{R}^{d\times d}$. In contrast to DE, it does not replace members on an individual basis, but uses a weighted subset of $M < N$ members with the lowest objective values to update the mean $\bm{\mu}$ and covariance matrix $\Sigma$ for the next iteration.

Downhill simplex (simplex) or Nelder-Mead search also updates a population of vectors in each iteration~\cite{nelder1965}. However, this population is a simplex of only $N=d+1$ vectors (i.e. a triangle in $\mathbb{R}^2$, a tetrahedron in $\mathbb{R}^3$ etc.) which is much smaller than typical populations of DE and CMA-ES. New candidates are created through processes called \textit{reflection}, \textit{expansion} and \textit{contraction} of the worst-performing vector of the simplex. If those three candidates do not lead to a specific improvement, the simplex shrinks towards its best performing vector $\bi{p}_{\rm best, simplex}$. Due to the small population, simplex is a more local minimization method that typically converges faster than other heuristic methods as in the worst case $d+2$ evaluations, or in the best case only one evaluation of the objective is required per iteration.

None of the above algorithms handle noise in an explicit manner. However, the population updates are often based on many vectors and corresponding objective evaluations (e.g. $N$ individual comparisons for DE, averaging over $M$ vectors for CMA-ES) such that the impact of noise is diminished. A different approach is taken in the LILDE extension of DE. This algorithm re-evaluates members that have survived a specific number of generations~\cite{Geisel2013}, further reducing the impact of noise by removing population members whose fitness value is only good due to noise. 
We speculate that PSO and simplex can be most severely misled by noise. For these methods a single vector and corresponding noisy evaluation can act as an attractor of the population ($\bi{p}_{\rm best, swarm}$ for PSO and $\bi{p}_{\rm best, simplex}$ for simplex). As long as no lower value is observed, these vectors are not updated. 

\subsection{Bayesian optimization}
\label{sec:BO}

BO is a sequential method that uses previous observations $\mathcal{D} = \{\tilde f(\bi{p}_1),\dots, \tilde f(\bi{p}_k)\}$ to train a stochastic machine learning model which is used to determine a new parameter vector $\bi{p}_{k+1}$ to evaluate~\cite{shahriari2015}. The stochastic model, a Gaussian process, can be regarded as a posterior distribution over random functions given the observations~\cite{rasmussen2006}. For each parameter vector $\bi{p}^*$ it predicts a normal distribution of possible function values $y(\bi{p}^*)\sim\mathcal{N}\left(\mu(\bi{p}^*),\sigma^2(\bi{p}^*)\right)$, where
\begin{eqnarray}
	\mu(\bi{p}^*) &=& \mu_0 + \sigma_0^2 \sum_{i,j=1}^k \kappa(\bi{p}^*, \bi{p}_i) \Sigma_{ij}^{-1} \left[f(\bi{p}_j) - \mu_0 \right]\,, \\
	\sigma^2(\bi{p}^*) &=& \sigma_0^2 - \sigma_0^4 \sum_{i,j=1}^k \kappa(\bi{p}^*, \bi{p}_i) \Sigma_{ij}^{-1} \kappa( \bi{p}_j, \bi{p}^*)\,.
\end{eqnarray}
Here, $\mu_0$ and $\sigma_0^2$ are hyperparameters for the mean and variance of the objective function. The covariance matrix $[\Sigma]_{ij} = \sigma_0^2 \kappa(\bi{p}_i, \bi{p}_j) + \eta^2\delta_{i j}$ depends on a positive definite covariance function $\kappa(\cdot,\cdot)$ and on an additional hyperparameter for the noise variance $\eta^2$. The best choice of the covariance function depends on the assumed differentiability of the objective $f(\bi{p})$~\cite{rasmussen2006}. In this work, the considered BO optimizers use the Mat\'ern-$5/2$ covariance function,
\begin{eqnarray}
	\kappa(\bi{p}_1, \bi{p}_2) &=& \left(1 + \sqrt{5} d + \frac{5}{3} d^2\right)\exp(-\sqrt{5} d) \\
	\text{with}\;\; d &=& \sqrt{\sum_{i=1}^d\frac{(p_{1,i}-p_{2,i})^2}{l_i^2}}\,.
\end{eqnarray}
The length scales $l_1,\dots,l_d$, at which the covariance decreases, are another set of hyperparameters. The value of all hyperparameters is determined by their maximum likelihood estimate~\cite{rasmussen2006}.

The next sampling point $\bi{p}_{k+1}$ is chosen by some infill criterion. A common choice is to maximise the expected improvement of the predicted objective value distribution
\begin{equation}
	{\rm EI}(\bi{p}^*) = {\rm E}\left[\max(0, y_{\rm min} - y(\bi{p}^*) )\right]
\end{equation}
with respect to the lowest seen objective value $y_{\rm min}$.

In the noiseless case ($\eta^2=0$), the predicted variance $\sigma^2(\bi{p}^*)$, and thus ${\rm EI}(\bi{p}^*)$, tends to zero if the number of neighbouring observations increases. Therefore, a local minimum will always be eventually escaped, since ${\rm EI}(\bi{p}^*)$ will be larger in parameter regions with less data and more predicted variance. In fact, it has been shown that the expected improvement strategy converges in the noiseless case at a near optimal rate to the \textit{global} minimum if the objective belongs to the reproducing kernel Hilbert space with kernel $\sigma_0^2 \kappa(\bi{p}_1, \bi{p}_2)$~\cite{bull2011}.

\subsection{Bayesian optimization of noisy functions}

In the noisy case, $\sigma^2(\bi{p}^*)$ and thus ${\rm EI}(\bi{p}^*)$ do not tend to zero, even if $\bi{p}^*$ is an observed point. This can lead to an oversampling of local minima. Moreover, the best observed value $y_{\rm min}$ might be corrupted by noise. Several extensions of BO have been proposed for noisy settings~\cite{picheny2013, letham2019}. In this work, we focus on the noisy expected improvement strategy (NEI)~\cite{letham2019}. For this, $F$ sets of random function values $\mathcal{F}_i=\{y_{i,1},\dots,y_{i,k}\}$, $i=1,\dots,F$ are drawn from the noisy Gaussian process posterior at the observed points $\bi{p}_1,\dots,\bi{p}_k$. The fantasies $\mathcal{F}_i$ are thus possible values of $f(\bi{p})$ that are compatible with the noisy observations $\mathcal{D}$. Each $\mathcal{F}_i$ is used to train the noiseless Gaussian process and to determine a noiseless expected improvement ${\rm EI}_i(\bi{p}^*)$ with respect to the corresponding minimum $y_{\rm min, i} = \min(\mathcal{F}_i)$. The noisy expected improvement is then defined as
the average over the fantasies
\begin{equation}
	{\rm NEI}(\bi{p}^*) = \frac{1}{F} \sum_{i=1}^F  {\rm EI}_i(\bi{p}^*)\,.
\end{equation}

\subsection{Different implementations of Bayesian optimization}
\label{sec:BO_implementation}
BO is a relatively complex method for which implementation details can have a large influence on the convergence and computational overhead.
Therefore, in this work three different implementations of BO are considered, the open-source package FMFN~\cite{FMFN}, the open-source AX framework~\cite{AX} provided by Meta Platforms Inc., and the BO optimizer (JCM) included in the commercial software JCMsuite, developed by some of the authors. The three methods follow, for example, different strategies to maximise acquisition functions (either ${\rm EI}(\bi{p}^*)$ or ${\rm NEI}(\bi{p}^*)$). 

The open-source package FMFN~\cite{FMFN} uses the best result from \num{10000} random samples of the acquisition function and 10 local L-BFGS-B optimizations started from random initial positions. Random sampling can result in samples of low quality for larger dimensionality $d$ of the parameter space. For example, for $d=18$ parameters, \num{10000} samples correspond to effectively about \num{1.7} samples per dimension which can be considered  extremely sparse.
The AX framework first evaluates the acquisition function at, by default, \num{1000} random positions. The best \num{20} positions are further optimized by the local L-BFGS-B or SLSQP method. The JCM optimizer performs an initial DE maximisation of the acquisition function. The obtained maximum and the previous sample with the lowest objective value are then tuned by a local L-BFGS-B optimization. 

AX and JCM can detect the noise level $\eta^2$ in order to perform a maximisation of ${\rm NEI}(\bi{p}^*)$. FMFN assumes by default that the objective function is noiseless. The same holds for the JCM optimizer with disabled noise detection.

Within our collaboration~\cite{optimal}, we developed different strategies with the goal of limiting the computational overhead of optimization based on noisy experimental data. For this work, the methods where implemented into the JCM optimizer. 
To compute ${\rm NEI}(\bi{p}^*)$ without requiring $F$ additional hyperparameter optimizations, the optimized length scale hyperparameters of the noisy Gaussian process are used as proposed in~\cite{letham2019}. Exploiting the fact that the noiseless covariance matrix $[\Sigma]_{ij}^{\rm noiseless} = \sigma_0^2 \kappa(\bi{p}_i, \bi{p}_i)$ is identical for each of the $F$ Gaussian processes up to different optimal prefactors $\sigma_0^2$, the expensive step of the matrix inversion (i.e. Cholesky decomposition) has to be done only once. Moreover, following the batch optimization strategy in~\cite{letham2019} the next sampling point is already computed during a pending evaluation of the objective. The numerical effort of this computation, e.g. the maximum number of DE iterations, is adapted to the runtime of the experiment to reduce the waiting time, in the ideal case, to zero~\cite{schneider2019}.

\section{Experimental set-up}\label{sec:setup}

\begin{figure}[ht]
    \centering
	\includegraphics[width=\textwidth]{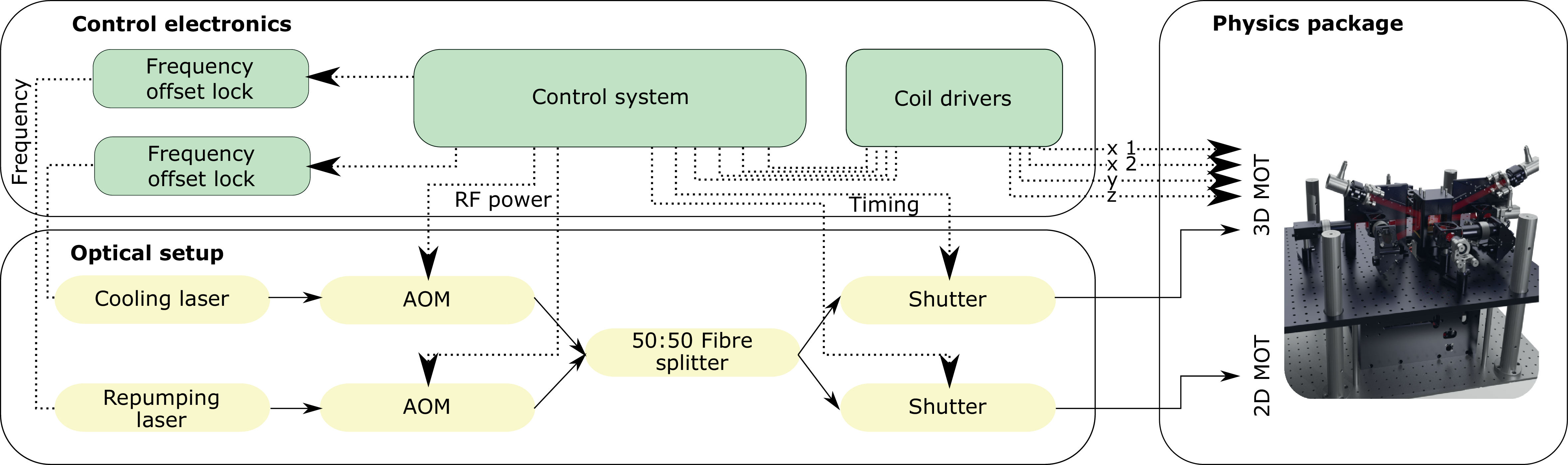}
	\caption{A block diagram showing the relevant parts of the experimental setup used within this paper, alongside a rendering of the apparatus (on the right). Dashed lines indicate electrical and solid lines optical signals, similarly green shading represents electronics components and yellow shading optical components.}
	\label{fig:ExpOverview}
\end{figure}

The apparatus used within this paper is described in Fig.~\ref{fig:ExpOverview}.
It is based on a commercial RuBECi vacuum system, alongside a 2D+ MOT platform, magnetic coils ~\cite{ColdQuanta2023}, and an optical system for the 3D MOT beams.
A rendering can be seen in the box labelled `physics package' in Fig.~\ref{fig:ExpOverview}.
The vacuum chamber consists of two glass cells separated by a differential pumping stage. 
One chamber is used to create a 2D+ MOT which acts as an atom source.
The second chamber acts as a `physics package', where atoms are trapped in a 3D-MOT before further cooling.
Here, two separately controllable coils, aligned along the x-axis, provide a magnetic gradient.
Additionally, two pairs of Helmholtz coils, aligned along the y- and z-axis of the experiment, are used to generate offset fields for cancelling stray magnetic fields. 
The laser light for trapping in the 3D-MOT is fed into the system via three collimators providing beams with a diameter of \SI{13.2}{\milli\meter} which are retro-reflected.
The atoms are imaged using absorption imaging. 

As shown in the `optical setup' box of Fig.~\ref{fig:ExpOverview}, cooling ($F=2\rightarrow F'=3$) and repumping ($F=1\rightarrow F'=2$) light is generated via two micro-integrated DFB-MOPA modules~\cite{Wicht2017}.
The lasers are frequency stabilized via offset locks to a reference laser locked to the $^{85}$Rb$~F=3\rightarrow F'=3/4$ crossover. 
The optical power delivered to the experiment is modulated and switched via acousto-optic modulators (AOM), then combined on a 50:50 fibre splitter, before additional extinction provided by shutters. No active intensity stabilization is present in the setup.

In short, the experimental sequence starts with the loading of the 3D-MOT by a 2D+ MOT, followed by the compression and molasses phases, and ends with the imaging of the atoms.
A selection of experimental parameters from this sequence will be used in order to perform the benchmarking. 

During the MOT phase, the detuning and optical power of the cooling and repumper lasers can be varied, as well as the currents delivered to both gradient coils and the z-offset coils.

Within the molasses phase, the power and detuning of cooling and repumper lasers can be ramped over a variable duration or switched, and the coil currents can also be modified as in the MOT phase.
These tunable molasses parameters are the start point of the ramp for the optical power of the repumper, the endpoints of the ramps for the optical power of cooler and repumper, the current applied to all coils responsible for generating offset fields, and the overall length of the molasses phase. Since the molasses phase is short compared to the rest of the sequence, this parameter does not significantly influence the run-time and thus the result of the comparison. This sequence is used for all measurements presented in the following benchmarking measurements. A list of the parameters including their limits can be found in ~\ref{tab:parameters}.

\begin{figure}[h]
	\centering
	\includegraphics[width=\textwidth]{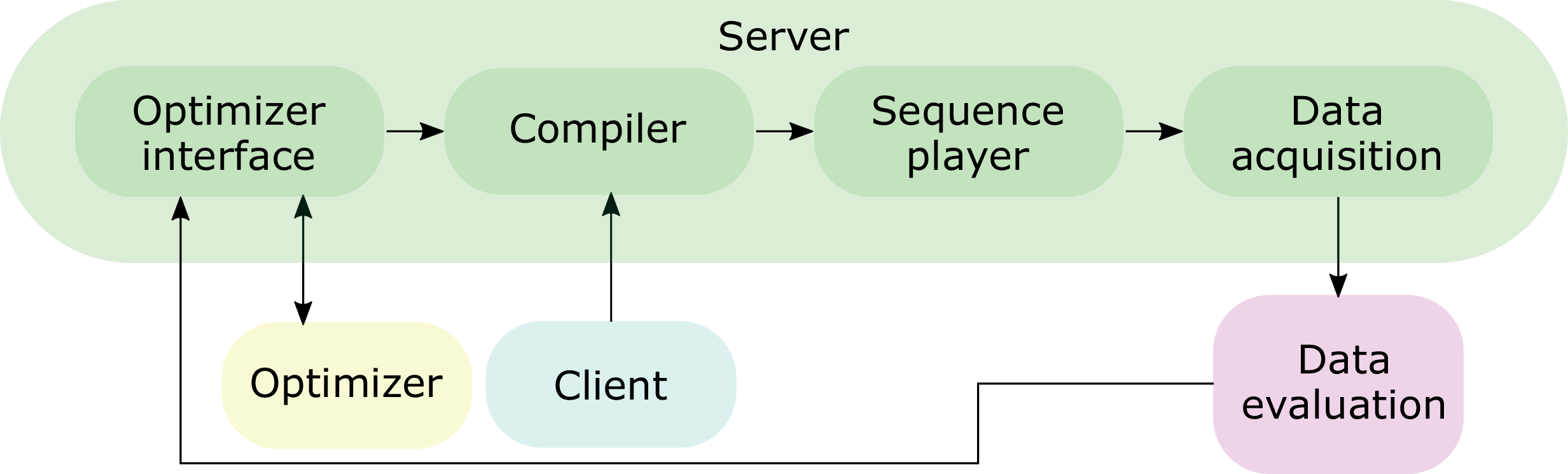}
	\caption{Schematic (block diagram) of the control software showing the server's workflow running the experiment with its sub-processes. Using the client, a sequence is created to be used by the server for compilation and later used by the sequence player. Feedback for the optimizer is provided using the acquired data via the data evaluation software. An adapted sequence is then fed into the compiler. }
	\label{fig:ControlChart}
\end{figure}

After a sequence is performed by the experiment, absorption images are evaluated and post-processed in tailored Python analysis code~\cite{Pical}.
Atom number and optical density can be evaluated via either a full 2D Gaussian fit, or via two 1D fits of row and column sums.
This information is then fed into the optimization algorithm which uses the result to generate a new sequence.
This is shown in Fig.~\ref{fig:ControlChart}.

The offset frequencies, optical powers, coil currents, and timings can be controlled via the Sinara system, a schematic of which is shown in the `Control electronics' box of Fig.~\ref{fig:ExpOverview}. 
The Sinara system is a Field Programmable Gate Array (FPGA) based real-time system, developed by Quartiq~\cite{M-labs2023}. 
The FPGA board is augmented with digital in/out breakout boards as well as DDS and DAC cards for controlling the experiment. 
By using these cards we control our frequency offset locks, coil drivers, shutters and AOMs.
The control software of the experiment is a server-client based system using Python.
It is described in \ref{App:Control}.

\section{Results}\label{sec:Results}

In the following sections we present benchmarking results of the performance of several optimizers using post-molasses atom number after \SI{10}{\milli\second} of free expansion as the test problem. 
The problem is chosen due to its resilience to false positives and the potential to adjust the noisiness of the data by worsening the imaging lasers frequency lock and reducing the rubidium gas pressure in the 2D+ MOT. 
The fitness function for this problem is $f= -N_{\rm atoms}$, which converts the maximisation of atom number into a minimization problem as required by the optimizers used. 
We also note that this experiment typically takes approximately \SI{2.5}{\second} per shot. 

In the 10-dimensional case we vary parameters which influence the number of trapped atoms, such as laser frequencies, optical powers and magnetic fields. 
A complete list of the parameters chosen including the limits can be found in the table in Appendix~\ref{app:parameters}.
The MOT loading time was excluded due to the effect it would have on the duration of the experiment and thus the comparability of the results. 
In addition, we included the start of the molasses phase: the frequencies of the cool and repumper lasers, as well as the starting point of the power ramp of the cooling laser. 

For the 18-dimensional case we extended this list of parameters to include all remaining variables influencing the molasses, namely the endpoint of the cooler power ramp, the start and end point of the repumper power ramp, the offset fields during molasses, and the temporal length of the molasses phase. 
Since the length of the molasses phase is short, including this as an optimization parameter does not noticeably affect the results of the comparison.
The power ramps are linear and frequency is instantaneously switched.

The optimizers were tested using the following procedure.
Each algorithm is run using the default settings for 400 iterations and repeated 10 times to obtain an average performance.
The choice of 400 iterations allows for many algorithms to reach a plateau whilst also limiting the time needed for data acquisition to a reasonable amount.
In order to remove systematics due to experimental drift, the algorithms are interleaved rather than run sequentially, that is, each algorithm is run once, one after another, and then this process repeated 10 times.
Parameter limits are chosen such as to reduce the size of the parameter space where no atoms are measured.

In the following figures, we show the evolution of the mean fitness (i.e. atom number) over time, with the shaded area representing the standard error of the 10 runs. 
The optimizers received no initial set and are initialized randomly within the defined parameter space.

As described in Sec.~\ref{sec:optimizers}, in order to provide comprehensive benchmarking, we compare at least one algorithm per family: CMA-ES ; the BO methods JCM with noise detection, FMFN and AX; PSO; variations on differential evolution LILDE and DE; RS; and simplex.
We benchmark for both 10- and 18- dimensional space as well as two different noise levels.

\begin{figure}[h]
	\centering
	\includegraphics[width=1\columnwidth]{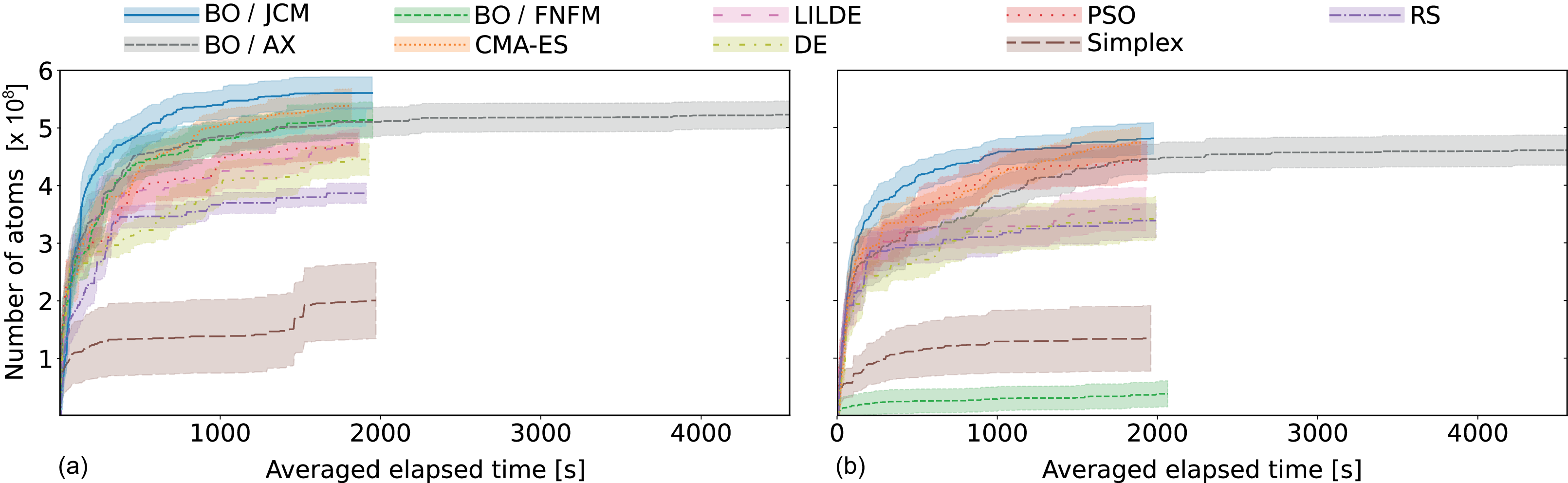}
	\caption{Development of the fitness function (maximization of atom number) over time for different optimizers for (a) a 10-dimensional parameter space and (b) a 18-dimensional parameter space. Each algorithm runs for 400 iterations. The lines represent the mean value, while the shaded areas show the standard deviation over 10 repetitions. }
	\label{fig:Comparission_10_Param}
\end{figure}

\subsection{10-dimensional optimization} \label{sec:10D}

Results for 10-dimensional optimization are shown in Fig.~\ref{fig:Comparission_10_Param}. 
RS is considered a baseline optimization that does not take the information from previous evaluations into account. 
In our benchmarking, it swiftly finds sets of parameters producing a signal, 
which indicates that a signal can be found throughout the majority of the parameter space. 
Although RS is considered a baseline method, we see that simplex consistently performs worse, reaching a maximum of \num{2.0(0.7)e8} atoms compared to \num{3.9(0.2)e8} atoms for RS, as well as taking longer to reach these values. 
As described in Sec.~\ref{sec:Heuristic}, simplex is a local minimization method and does not incorporate strategies to limit effects of noise.
Noisy observations can deteriorate the algorithm by attracting population members to regions with relatively low performance.

Following simplex and RS in performance are DE, LILDE and PSO.  
All three reach similar fitness values (\num{4.5(0.3)e8}, \num{4.7(0.2)e8}, and \num{4.7(0.2)e8} atoms respectively) but do not plateau within the given number of iterations.
The graphs show jumps in performance at certain times for all three. 
This is most visible in the LILDE algorithm at \SI{400}{\second} and \SI{1000}{\second}. 
The jumps are due to these optimizers relying on a generational approach where gained knowledge is collected and combined after a defined number of iterations to generate new sets. 
The updated process produces a sharp increase in the reached fitness value.
This indicates that an adjustment of the default generation size may increase the performance.

Bayesian optimizers FMFN and AX lie in third and fourth place. 
Both perform surprisingly equally in terms of the reached fitness with \num{5.1(0.3)e8} and \num{5.2(0.2)e8} atoms respectively.
In terms of optimization speed, they perform similarly to begin with, however, over the full iteration number a clear advantage of FMFN over AX is visible. We attribute the significant computational overhead of AX to the more elaborate computation of expected improvement which requires predictions from $F$ (default $F=20$) Gaussian processes instead of only one.
Based on the plateauing of fitness it seems that neither optimizer would significantly benefit from a higher number of iterations. 

CMA-ES performs second best. 
Here we see a fitness of \num{5.4(0.3)e8} atoms which is close to the best reachable value.
Upon reaching \num{400} iterations, the fitness is still increasing.
As a result, it is possible that a higher fitness value could be reached with further iterations and it may perform similar to the best algorithm given enough time.
It is surprising that such a simple optimizer still yields such good performance compared to more elaborate approaches. 
As such, we conclude that the described approach of updating the mean vector from a set of points is robust against the noise levels in this problem.  

The best performing optimizer in our list is the JCM Bayesian optimizer, which was extended for this work. 
It outperforms the others from start to end.
The computation time is short compared to AX due to strategies like precomputing next samples and sharing information between the $F$ Gaussian process regressions (see the theory section~\ref{sec:BO_implementation}).
It reaches the best fitness (\num{5.6(0.3)e8} atoms) after $\approx$\SI{1500}{\second} reaching a plateau afterwards. The parameters used to achieve the best fitness are shown for the 4 best performing optimizers in App.\ref{app:Input10parameters}.

\subsection{18-dimensional optimization}
In the 18-dimensional case we find that all optimizers perform consistently worse compared to the 10-dimensional case. 
Since all limits stayed the same 
as those used in the 10-dimensional case, one would expect the optimizers to reach the same, or possibly better, fitness as before. We assume that the larger parameter space makes the optimization problem much harder such that only lower fitness values are found within the same number of iterations.
However, we note that we are unable to rule out systematic drifts in the experiment between the time the 10- and 18-dimensional data was taken.

When we look at each optimizer individually, we find that FMFN is now the worst performing optimizer, notably worse even than simplex and RS. We attribute this behaviour to its very sparse search for good samples in higher dimensional problems, as described in Sec.~\ref{sec:BO_implementation}.

LILDE and DE now barely perform better than RS. By default, in these algorithms, the population size is \num{15} times the number of dimensions. Therefore, for $d=18$ dimensions, there are \num{270} evaluations per iteration meaning that these optimization approaches reduce to effectively random sampling within 400 iterations.

As in the 10-dimensional example, the high overhead of AX results in it taking far longer than any other optimizer to finish 400 iterations. Despite the higher computational effort and thus long optimization times, AX  still performs well for higher dimensional problems, reaching a plateau of \num{4.6(0.3)e8} atoms.

PSO performs similarly to AX in terms of fitness (reaching \num{4.4(0.3)e8} atoms) but takes much less time. 

The best performing optimizers are again CMA-ES and the JCM optimizer reaching similar fitness values of \num{4.7(0.3)e8} and \num{4.8(0.3)e8} atoms respectively.
CMA-ES takes marginally less time for \num{400} iterations, and does not reach a plateau value, therefore its fitness may eventually exceed the JCM optimizer.
Despite this, the JCM optimizer reaches a close-to-optimal fitness value far more quickly than any other optimizer.

In conclusion, one can see the commercial Bayesian optimizer, JCM, performing best, closely followed by CMA-ES representing the best open-source optimizer available for this use case. 
Other optimizers perform consistently worse and degrade in performance, when compared to the lower dimensional problem. 
We see this most prominently for FMFN. The parameters used to achieve the best performing sequences are shown for the best four optimizers in the table in App.~\ref{app:Input18parameters}.

\subsection{Influence of noise on the experiment optimization}\label{sec:noise}

In order to explore the robustness of the optimizers to noise, we tested the algorithms at two different noise levels for the 10-dimensional problem described above. 
The different noise levels were achieved in two ways: by changing the dispenser current in the 2D+ MOT we can reduce the number of atoms loaded into the 3D MOT; and by changing the PID parameters of the imaging laser we can introduce instabilities in its frequency and thus in the number of atoms measured. 
The lower atom number reached in the higher noise test case is due to the experimental conditions (i.e. a reduced loading rate) rather than the performance of the optimizers.
The two noise levels have a fluctuation in atom number of \SI{\pm5.6}{\percent} and \SI{\pm19.5}{\percent}, where we note that the lower noise level is the typical performance of our experiment.
We compared the highest performing optimizers, namely AX, FMFN, CMA-ES and JCM, alongside the JCM optimizer without its noise detection feature.

\begin{figure}[h]
    \centering 
    \includegraphics[width=\textwidth]{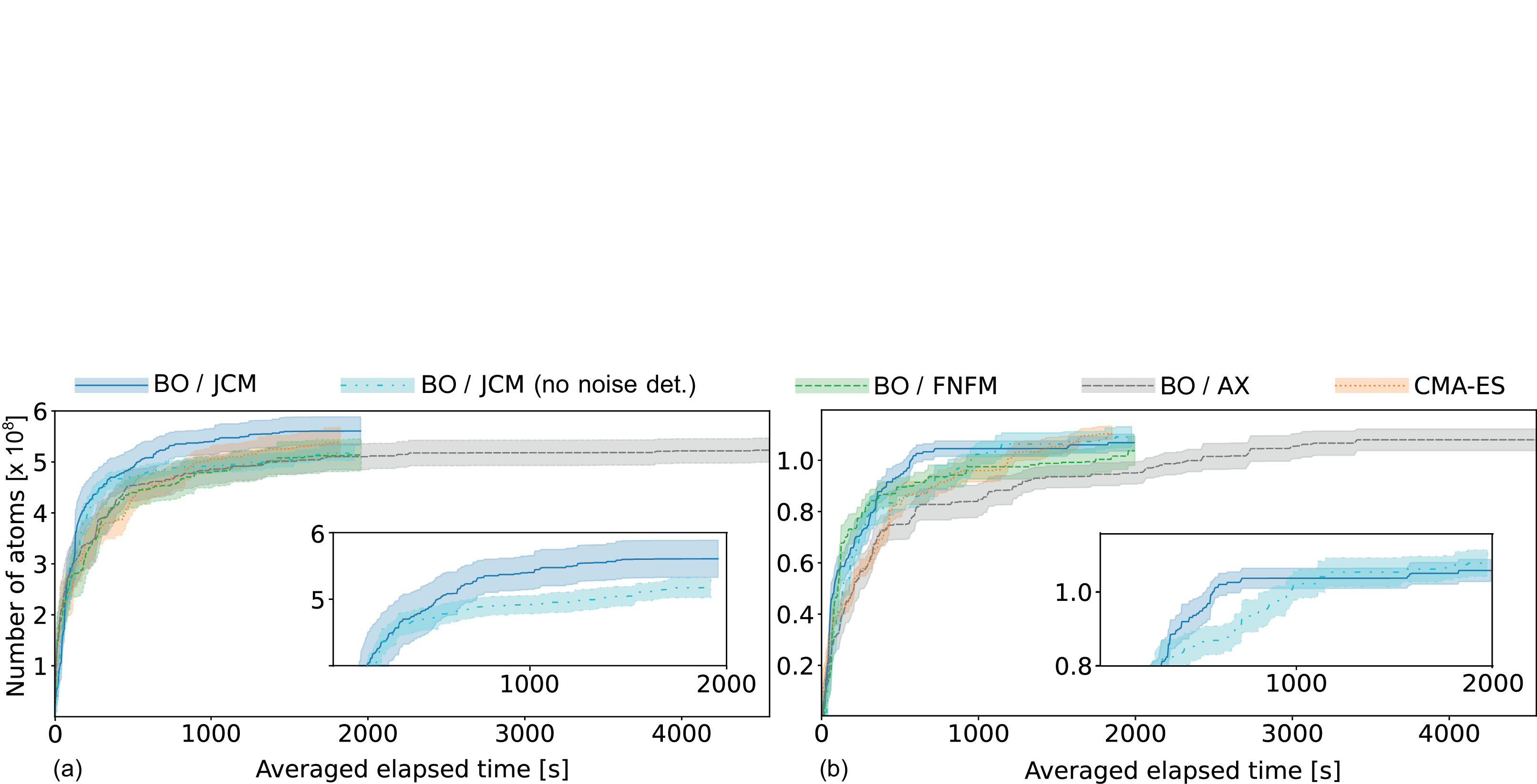}
	\caption{Development of the fitness function (maximization of atom number) over time for the highest performing optimizers with different noise levels. Figure (a) shows optimization with an atom number noise level of~\SI{\pm5.6}{\percent}, while (b) shows the same for a noise level of~\SI{\pm19.5}{\percent}. The JCM optimizer is tested with and without noise detection activated, with this comparison highlighted in the insets. The reduced atom number reached in (b) is due to the experimental conditions choosen to increase the noise level. Each algorithm runs for 400 iterations. The lines represent the mean value, while the shaded area shows the standard deviation over 10 repetitions. }
	\label{fig:Noise}
\end{figure}

In the low noise case (Fig.~\ref{fig:Noise}), the same trends are visible as discussed in Sec.~\ref{sec:10D}.
However, by comparing the JCM algorithm with and without noise detection we can see that the additional feature does not increase early optimization speeds, instead it results in a slightly higher fitness value of \num{5.6(0.3)e8} (with noise detection) compared to \num{5.2(0.2)e8} (without noise detection) atoms and in that higher fitness being reached faster.

In the high noise case, all five algorithms tested perform relatively similarly in terms of fitness.
FMFN reaches the lowest fitness value (\num{1.03(0.06)e8} in \SI{2000}{\second}), followed by AX at \num{1.08(0.04)e8} in \SI{4800}{\second}, JCM with and without noise detection reach \num{1.07(0.03)e8} and \num{1.09(0.04)e8} atoms in \SI{2000}{\second}, and finally CMA-ES reaches a fitness value of \num{1.10(0.03)e8} atoms in \SI{1900}{\second}.
Although we have ranked the algorithms here according to the mean fitness value, the differences between them are almost all within error margins.
The main differences in the algorithms can actually be seen in the speed of optimization.
For example, AX is the slowest optimizer at all points in the process, and FMFN slows dramatically after performing fastest in very early stages.

Due to the additional noise detection capabilities of the algorithm developed here, it was expected to perform the best in a high noise environment.
The benchmarking data shows that it does indeed optimize faster than all other optimizers and to a higher fitness than all optimizers except CMA-ES.
In fact, it reaches close to its final value within \SI{720}{\second}.
This is roughly half the time taken for CMA-ES to reach the same value.
We assume that CMA-ES performs so well in terms of final fitness due to averaging over many vectors which limits the effect of noise (Sec.~\ref{sec:Heuristic}). 
Nevertheless, the noise detection function provided by JCM significantly improves the speed of optimization compared to the algorithm without it, and enables it to perform better than the other BOs tested.
In general we find that most optimizers do not reach a plateau in the available iterations. 

In summary, the influence of noise extends the iterations and thus time needed for the optimization. 
It also generally reduces the initial speed of optimization considerably, which means that more time is required to achieve a `good enough' value. 
There is a trade-off in performance between speed and final fitness, with CMA-ES providing a marginally better final fitness, and the JCM with noise detection algorithm reaching a high fitness value much quicker than other algorithms.

\section{Outlook} \label{sec:outlook}

In conclusion, we tested different optimizers and showed their performance for different dimensionalities. Additionally, the developed implementation of JCM's efficient noise detection feature was benchmarked against the three best performing algorithms at two different noise levels. 
The best performing optimizers were found to be JCM (with noise detection) and the open source CMA-ES, with JCM providing faster optimization.

Optimization of the problem using 18 dimensions rather than 10 resulted in a worse performance for all optimizers. This shows indications that optimizing more parameters at once is not always better when it comes to the optimization of real world experiments.
We also tested the influence of noise on optimizer performance, finding that with increased noise AX, JCM, FMFN and CMA-ES reach similar fitness, however, JCM and CMA-ES continued to offer the best performance in terms of speed and fitness. 
We were able to confirm that the noisy expected improvement strategy implemented into the JCM optimizer for this work improves the speed of optimization and in some cases also the final fitness value.

When choosing a suitable optimizer for an experiment, one must make a trade-off between speed, final fitness, and cost. 
The open source algorithm CMA-ES provided a marginally better final fitness, whereas the JCM optimizer with noise detection reached a high fitness value much quicker than other algorithms. 
Therefore, CMA-ES would be better suited to an experiment which would be optimized once or infrequently.
However, for experiments requiring very regular optimization, such as mobile experiments, the speed benefits of the JCM optimizer would enable more efficient operation.

The results presented could be extended further via investigations at longer iteration numbers or in more noisy conditions.

\section{Acknowledgements}
\noindent This work is supported by the German Space Agency
(DLR) with funds provided by the BMWK under grant numbers No.~ 50WM2067, 50WM2175, 50WM2055, as well as the German Federal Ministry of Education and Research (BMBF Forschungscampus MODAL, project number 05M20ZBM).

\clearpage

\appendix
\setcounter{section}{0}
\section{Experimental parameters} \label{app:parameters}
\begin{table}[h]
	\caption{\label{tab:parameters}Optimization parameters and their limits. For the 10 parameter optimization, the parameters up to the line were used, for 18 parameter optimization, all parameters were used. The optimizer values are the values used by our control system to run the experiment. The physical frequencies are quoted as offsets from the ($F=2\rightarrow F'=3$) and repumping ($F=1\rightarrow F'=2$) transition respectively. The conversion factors of the voltage to magnetic fields are not given for the gradient coils. They can generate offset fields as well as gradient fields depending on the voltage of the other gradient coils so no limit independent of the setpoint of the other coil can be given.}
	\begin{indented} 
		\lineup
		\item[]\begin{tabular}{@{}llllll}
			\br
			 Stage & Parameter & \multicolumn{2}{c}{Optimizer values} &\multicolumn{2}{c}{Physical quantities}\\
			&& Lower & Upper & Lower & Upper \\
			\mr
			MOT & Cool - frequency	& \SI{143.0}{\mega\hertz} & \SI{146.5}{\mega\hertz} &\SI{2.16}{\mega\hertz}&\SI{-25.84}{\mega\hertz}\\
			MOT & Repumper - frequency&\SI{317}{\mega\hertz} & \SI{323}{\mega\hertz} &\SI{-32.3}{\mega\hertz}&\SI{69.7}{\mega\hertz}\\
			MOT & Cool - power		& \SI{0}{\decibel} & \SI{20}{\decibel} & \SI{30}{\milli\watt} & \SI{4}{\milli\watt}\\
			MOT & Repumper - power	& \SI{0}{\decibel} & \SI{30}{\decibel} & \SI{3.4}{\milli\watt} & \SI{0.1}{\milli\watt}\\
			MOT & Gradient coils - X1	& \SI{-4}{\volt} & \SI{-2.5}{\volt} & - & - \\
			MOT & Gradient coils - X2	& \SI{2.5}{\volt} & \SI{4}{\volt} & - & - \\
			MOT & Offset coil - Z &  \SI{0.6}{\volt} & \SI{1.0}{\volt} & \SI{1.4}{\gauss} & \SI{2.3}{\gauss} \\ 
			Molasses & Cool - frequency& \SI{165}{\mega\hertz} & \SI{175}{\mega\hertz} &\SI{-174}{\mega\hertz}&\SI{254}{\mega\hertz}\\ 
			Molasses & Repumper - frequency& \SI{310}{\mega\hertz} & \SI{317}{\mega\hertz} &\SI{-152}{\mega\hertz}&\SI{-33}{\mega\hertz}\\ 
			Molasses & Cool - power start& \SI{0}{\decibel} & \SI{20}{\decibel} & \SI{30}{\milli\watt} & \SI{4}{\milli\watt}\\
			\mr
			Molasses & Repumper - power start& \SI{10}{\decibel} & \SI{20}{\decibel} & \SI{1.9}{\milli\watt} & \SI{0.2}{\milli\watt}\\
			Molasses & Cool - power end& \SI{18}{\decibel} & \SI{30}{\decibel} & \SI{5}{\milli\watt} & \SI{0.1}{\milli\watt}\\
			Molasses & Repumper - power end& \SI{18}{\decibel} & \SI{30}{\decibel} & \SI{0.4}{\milli\watt} & \SI{0}{\milli\watt}\\
			Molasses & Gradient coils - X1	& \SI{-0.1}{\volt} & \SI{0.1}{\volt} & - & - \\
			Molasses & Gradient coils - X2	& \SI{0}{\volt} & \SI{0.2}{\volt} & - & - \\
			Molasses & Offset coil - Z &  \SI{-0.3}{\volt} & \SI{0}{\volt} & \SI{-6.8}{\gauss} & \SI{0.0}{\gauss} \\
			Molasses & Offset coil - Y &  \SI{-0.1}{\volt} & \SI{0.1}{\volt} & \SI{-6.8}{\gauss} & \SI{6.8}{\gauss} \\
			Molasses & Duration	& \SI{10}{\milli\second} & \SI{30}{\milli\second} & \SI{10}{\milli\second} & \SI{30}{\milli\second} \\
			\br
		\end{tabular}
	\end{indented}
\end{table}

\section{Best parameters reached in the 10 dimensional case} \label{app:Input10parameters}

    \begin{table}[H]
    	\caption{\label{tab:final_parameters}Mean values of the input parameters, including their standard deviation, used to reach the best fitness. The parameters of the four best performing optimizers are shown. A big standard deviation shows either little influence of this parameter on the result or the optimizer having difficulties in finding the optimal set.}
   	\begin{indented} 
    		\lineup
    		\item[]\begin{tabular}{@{}llllll}
                \br
            Stage & Parameter & BO / JCM & CMA-ES & FMFN & AX \\
            \mr
            MOT & Cool - frequency [\si{\mega\hertz}]	& \num{145.08 \pm 0.08}{} &	\num{145.09 \pm0.07}{} &	\num{145.0 \pm0.1}{} &	\num{145.0 \pm0.1}{} \\
            MOT & Repumper - frequency [\si{\mega\hertz}] & \num{319.1\pm0.4}{}	& \num{319.0\pm0.2}{}	& \num{319.0\pm0.4}{} &	\num{318.9\pm0.3}{}	\\
            MOT & Cool - power [\si{\decibel}] & \num{5\pm5}{}	& \num{6\pm3}{}	& \num{7\pm2}{}	& \num{6\pm2}{}	\\
            MOT & Repumper - power [\si{\decibel}] & \num{2\pm2}{}	& \num{4\pm2}{}	& \num{8\pm5}{}	&\num{4\pm2}{}	\\
            MOT & Gradient coils - X1 [\si{\volt}] & \num{-3.72\pm0.08}{}	& \num{-3.80\pm0.08}{}	& \num{-3.8\pm0.2}{}	& \num{-3.7\pm0.1}{}	\\
            MOT & Gradient coils - X2 [\si{\volt}] & \num{3.96\pm0.05}{}	& \num{3.98\pm0.02}{}	& \num{3.97\pm0.05}{} & \num{3.9\pm0.2}{} \\
            MOT & Offset coil - Z [\si{\volt}] &  \num{0.82\pm0.08}{}	& \num{0.80\pm0.09}{}	& \num{0.8\pm0.1}{}	& 0\num{.82\pm0.04}{}	\\
            Molasses & Cool - frequency [\si{\mega\hertz}] & \num{172\pm2}{}	& \num{171\pm2}{}	& \num{171\pm2}{}	& \num{171\pm2}{}	\\
            Molasses & Repumper - frequency [\si{\mega\hertz}] & \num{312\pm2}{} & \num{313\pm3}{}	& \num{313\pm2}{}	& \num{312\pm3}{}	\\
            Molasses & Cool - power start [\si{\decibel}] & \num{10\pm3}{} & \num{15\pm4}{} & \num{11\pm4}{} & \num{10\pm6}{} \\

            \br
            
    		\end{tabular}
    	\end{indented}
    \end{table}

\section{Best parameters reached in the 18 dimensional case} \label{app:Input18parameters}

    \begin{table}[H]
    	\caption{\label{tab:final_parameters_18dim}Mean values of the input parameters, including their standard deviation, used to reach the best fitness. The parameters of the four best performing optimizers are shown. A big standard deviation shows either little influence of this parameter on the result or the optimizer having difficulties in finding the optimal set.}
   	\begin{indented} 
    		\lineup
    		\item[]\begin{tabular}{@{}llllll}
            \br
            Stage & Parameter & BO / JCM & CMA-ES & PSO & AX \\
            \mr
            MOT & Cool - frequency [\si{\mega\hertz}]	& \num{145.0\pm0.1}{} &	\num{145.0\pm0.1}{} &	\num{145\pm0.2}{} &	\num{145\pm0.1}{}\\
            MOT & Repumper - frequency [\si{\mega\hertz}] & \num{319.1\pm0.4}{}	& \num{319.2\pm0.4}{} & \num{318.5\pm0.6}{} &	\num{319\pm0.2}{}	\\
            MOT & Cool - power [\si{\decibel}] & \num{6\pm3}{}	& \num{7\pm2}{}	& \num{6\pm5}{}	& \num{7\pm2}{}	\\
            MOT & Repumper - power [\si{\decibel}] & \num{5\pm2}{}	& \num{6\pm3}{}	& \num{3\pm3}{}	&\num{6\pm2}{}	\\
            MOT & Gradient coils - X1 [\si{\volt}] & \num{-3.7\pm0.2}{}	& \num{-3.8\pm0.2}{}	& \num{-3.8\pm0.3}{}	& \num{-3.5\pm0.2}{}	\\
            MOT & Gradient coils - X2 [\si{\volt}] & \num{3.90\pm0.07}{}	& \num{3.95\pm0.01}{} & \num{3.9\pm0.2}{} & \num{3.6\pm0.2}{} \\
            MOT & Offset coil - Z [\si{\volt}] &  \num{0.9\pm0.1}{} & \num{0.81\pm0.09}{} & \num{0.8\pm0.1}{}	& \num{0.80\pm0.06}{}	\\
            Molasses & Cool - frequency [\si{\mega\hertz}] & \num{170\pm2}{}	& \num{170\pm2}{} & \num{169\pm3}{}	& \num{170\pm1}{}	\\
            Molasses & Repumper - frequency [\si{\mega\hertz}] & \num{312\pm1}{} & \num{312\pm2}{}	& \num{313\pm3}{}	& \num{313\pm1}{}	\\
            Molasses & Cool - power start [\si{\decibel}] & \num{9\pm5}{}  & \num{9\pm5}{} & \num{11\pm8}{} & \num{11\pm4}{} \\
            Molasses & Repumper - power start [\si{\decibel}] & \num{15\pm2}{} & \num{16\pm3}{} & \num{15\pm15}{} & \num{16\pm1}{} \\
			Molasses & Cool - power end [\si{\decibel}] & \num{24\pm2}{} & \num{23\pm4}{} & \num{24\pm4}{} & \num{23\pm2}{} \\
			Molasses & Repumper - power end [\si{\decibel}] & \num{22\pm2}{} & \num{22\pm3}{} & \num{24\pm5}{} & \num{23\pm1}{} \\
			Molasses & Gradient coils - X1 [\si{\volt}]	& \num{0.02\pm0.05}{} & \num{0.04\pm0.04}{} & \num{-0.03\pm0.06}{} & \num{0.00\pm0.03}{} \\
			Molasses & Gradient coils - X2 [\si{\volt}]	&\num{0.08\pm0.05}{} & \num{0.13\pm0.05}{} & \num{0.11\pm0.08}{} & \num{0.09\pm0.03}{} \\
			Molasses & Offset coil - Z [\si{\volt}] &\num{-0.18\pm0.07}{} & \num{-0.19\pm0.08}{} & \num{-0.2\pm0.1}{} & \num{-0.16\pm0.03}{} \\
			Molasses & Offset coil - Y [\si{\volt}] & \num{-0.01\pm0.03}{} & \num{0.02\pm0.02}{} & \num{0.02\pm0.07}{} & \num{0.00\pm0.02}{} \\
            Molasses & Duration [\si{\milli\second}] & \num{18\pm4}{} & \num{18\pm5}{} & \num{14\pm6}{} & \num{20\pm3}{} \\            
            \br
    		\end{tabular}
    	\end{indented}
    \end{table}

\section{Experimental control software}
\label{App:Control}

The control software of the experiment is a server-client-based system using Python.
In our system the client handles the graphical user interface and user-input to the experiment.


The server runs on the experiment control PC. 
It connects to all devices needed to run the experiment like cameras as well as Sinara, handles the communication between them and external software for data evaluation, storage and optimization. 
The workflow within the software for running the experiment can be seen in the server box of Fig.~\ref{fig:ControlChart}. 
Once the server receives instructions for a sequence, this table is compiled into instructions executable by the Sinara system before being queued for execution in the sequence player. 
After the sequence is played, the Sinara system signals to the server that data can be read out from devices. 
The acquired data is sent for evaluation in external software and the results are sent back to the server. 
Back at the server the data can be used for optimization processes which generate a new sequence using the suggestions from the optimization algorithm selected.

A single sequence can consist of multiple time steps addressing many channels of the Sinara hardware. 
The resulting table can become rather complex and thus laboratory routines can be optimized by the saving and loading of previously played sequences. 
Here, each unique, played sequence is stored in a MySQL database with a hash generated from its content for identification and to avoid duplication. 
Similarly, each time a sequence is played, a measurement hash is created, which includes the time of execution and any volatile settings like calibrations and camera settings. 
The measurement hashes are linked to the sequence hash and can be used to identify saved data.

Absorption images are evaluated and post-processed in tailored Python analysis software we call Pical (Picture analysis for cold atoms).
Atom number and optical density can be evaluated via either a full 2D Gaussian fit, or via two 1D fits of row and column sums.
The software is also capable of analysing sequences of pictures, which allows for the determination of the expansion rate or temperature of the cloud.
One can implement a range of other post-processing calculations such as calculating the phase-space density of the cloud, or by evaluating an alternative cost function.

\section*{References}
\bibliographystyle{iopart-num} 
\bibliography{Bib_test} 

\providecommand{\newblock}{}
\begin{thebibliography}{10}
\expandafter\ifx\csname url\endcsname\relax
  \def\url#1{{\tt #1}}\fi
\expandafter\ifx\csname urlprefix\endcsname\relax\def\urlprefix{URL }\fi
\providecommand{\eprint}[2][]{\url{#2}}

\bibitem{Dalibard1989}
Dalibard J and Cohen-Tannoudji C 1989 {\em Journal of the Optical Society of America B\/} {\bf 6} 2023 ISSN 0740-3224

\bibitem{metcalf1999}
Metcalf H~J and Van~der Straten P 1999 {\em Laser cooling and trapping\/} (Springer New York, NY)

\bibitem{Ketterle1996}
Ketterle W and Druten N~V 1996 {Evaporative Cooling of Trapped Atoms} {\em Advances In Atomic, Molecular, and Optical Physics\/} vol~37 (Elsevier) pp 181--236 ISBN 9780120038374 \urlprefix\url{http://www.sciencedirect.com/science/article/pii/S1049250X08601019 http://linkinghub.elsevier.com/retrieve/pii/S1049250X08601019}

\bibitem{Ketterle1999}
Ketterle W, Durfee D~S and Stamper-Kurn D~M 1999 Making, probing and understanding bose-einstein condensates (\textit{Preprint} \eprint{cond-mat/9904034})

\bibitem{Freier2016}
Freier C, Hauth M, Schkolnik V, Leykauf B, Schilling M, Wziontek H, Scherneck H~G, M{\"{u}}ller J and Peters A 2016 {\em Journal of Physics: Conference Series\/} {\bf 723} ISSN 17426596 (\textit{Preprint} \eprint{1512.05660})

\bibitem{Stray2022}
Stray B, Lamb A, Kaushik A, Vovrosh J, Rodgers A, Winch J, Hayati F, Boddice D, Stabrawa A, Niggebaum A, Langlois M, Lien Y~H, Lellouch S, Roshanmanesh S, Ridley K, de~Villiers G, Brown G, Cross T, Tuckwell G, Faramarzi A, Metje N, Bongs K and Holynski M 2022 {\em Nature\/} {\bf 602} 590--594 ISSN 14764687

\bibitem{Wu2019}
Wu X, Pagel Z, Malek B~S, Nguyen T~H, Zi F, Scheirer D~S and M{\"{u}}ller H 2019 {\em Science Advances\/} {\bf 5} 1--10 ISSN 23752548 (\textit{Preprint} \eprint{1904.09084})

\bibitem{menoret2018}
M{\'e}noret V, Vermeulen P, Le~Moigne N, Bonvalot S, Bouyer P, Landragin A and Desruelle B 2018 {\em Scientific reports\/} {\bf 8} 12300

\bibitem{antoni2022}
Antoni-Micollier L, Carbone D, M{\'e}noret V, Lautier-Gaud J, King T, Greco F, Messina A, Contrafatto D and Desruelle B 2022 {\em Geophysical Research Letters\/} {\bf 49} e2022GL097814

\bibitem{poli2014}
Poli N, Schioppo M, Vogt S, Falke S, Sterr U, Lisdat C and Tino G 2014 {\em Applied Physics B\/} {\bf 117} 1107--1116

\bibitem{Grotti2018}
Grotti J, Koller S, Vogt S, H{\"{a}}fner S, Sterr U, Lisdat C, Denker H, Voigt C, Timmen L, Rolland A, Baynes F~N, Margolis H~S, Zampaolo M, Thoumany P, Pizzocaro M, Rauf B, Bregolin F, Tampellini A, Barbieri P, Zucco M, Costanzo G~A, Clivati C, Levi F and Calonico D 2018 {\em Nature Physics\/} {\bf 14} 437--441 ISSN 17452481 (\textit{Preprint} \eprint{1705.04089}) \urlprefix\url{http://dx.doi.org/10.1038/s41567-017-0042-3}

\bibitem{takamoto2020}
Takamoto M, Ushijima I, Ohmae N, Yahagi T, Kokado K, Shinkai H and Katori H 2020 {\em Nature Photonics\/} {\bf 14} 411--415

\bibitem{gellesch2020}
Gellesch M, Jones J, Barron R, Singh A, Sun Q, Bongs K and Singh Y 2020 {\em Advanced Optical Technologies\/} {\bf 9} 313--325

\bibitem{boulder2021}
Collaboration B~A~C~O~N~B 2021 {\em Nature\/} {\bf 591} 564--569

\bibitem{elliott2018}
Elliott E~R, Krutzik M~C, Williams J~R, Thompson R~J and Aveline D~C 2018 {\em npj Microgravity\/} {\bf 4} 16

\bibitem{Frye2019}
Frye K, Abend S, Bartosch W, Bawamia A, Becker D, Blume H, Braxmaier C, Chiow S~W, Efremov M~A, Ertmer W, Fierlinger P, Franz T, Gaaloul N, Grosse J, Grzeschik C, Hellmig O, Henderson V~A, Herr W, Israelsson U, Kohel J, Krutzik M, K{\"{u}}rbis C, L{\"{a}}mmerzahl C, List M, L{\"{u}}dtke D, Lundblad N, Marburger J~P, Meister M, Mihm M, M{\"{u}}ller H, M{\"{u}}ntinga H, Nepal A~M, Oberschulte T, Papakonstantinou A, Perov{\v{s}}ek J, Peters A, Prat A, Rasel E~M, Roura A, Sbroscia M, Schleich W~P, Schubert C, Seidel S~T, Sommer J, Spindeldreier C, Stamper-Kurn D, Stuhl B~K, Warner M, Wendrich T, Wenzlawski A, Wicht A, Windpassinger P, Yu N and W{\"{o}}rner L 2021 {\em EPJ Quantum Technology\/} {\bf 8} 1 ISSN 2662-4400 (\textit{Preprint} \eprint{1912.04849}) \urlprefix\url{http://arxiv.org/abs/1912.04849 https://epjquantumtechnology.springeropen.com/articles/10.1140/epjqt/s40507-020-00090-8}

\bibitem{devani2020}
Devani D, Maddox S, Renshaw R, Cox N, Sweeney H, Cross T, Holynski M, Nolli R, Winch J, Bongs K {\em et~al.\/} 2020 {\em CEAS Space Journal\/} {\bf 12} 539--549

\bibitem{aveline2020}
Aveline D~C, Williams J~R, Elliott E~R, Dutenhoffer C, Kellogg J~R, Kohel J~M, Lay N~E, Oudrhiri K, Shotwell R~F, Yu N {\em et~al.\/} 2020 {\em Nature\/} {\bf 582} 193--197

\bibitem{Sidhu2021}
Sidhu J~S, Joshi S~K, G\"undo\u{g}an M, Brougham T, Lowndes D, Mazzarella L, Krutzik M, Mohapatra S, Dequal D, Vallone G, Villoresi P, Ling A, Jennewein T, Mohageg M, Rarity J~G, Fuentes I, Pirandola S and Oi D~K~L 2021 {\em IET Quantum Communication\/} {\bf 2} 182--217

\bibitem{Ahlers2022}
Ahlers H, Badurina L, Bassi A, Battelier B, Beaufils Q, Bongs K, Bouyer P, Braxmaier C, Buchmueller O, Carlesso M, Charron E, Chiofalo M~L, Corgier R, Donadi S, Droz F, Ecoffet R, Ellis J, Est{\`{e}}ve F, Gaaloul N, Gerardi D, Giese E, Grosse J, Hees A, Hensel T, Herr W, Jetzer P, Kleinsteinberg G, Klempt C, Lecomte S, Lopes L, Loriani S, M{\'{e}}tris G, Martin T, Mart{\'{i}}n V, M{\"{u}}ller G, Nofrarias M, Santos F~P~D, Rasel E~M, Robert A, Saks N, Salter M, Schlippert D, Schubert C, Schuldt T, Sopuerta C~F, Struckmann C, Tino G~M, Valenzuela T, von Klitzing W, W{\"{o}}rner L, Wolf P, Yu N and Zelan M 2022  (\textit{Preprint} \eprint{2211.15412}) \urlprefix\url{http://arxiv.org/abs/2211.15412}

\bibitem{Alonso2022}
Alonso I, Alpigiani C, Altschul B, Ara{\'u}jo H, Arduini G, Arlt J, Badurina L, Bala{\v{z}} A, Bandarupally S, Barish B~C, Barone M, Barsanti M, Bass S, Bassi A, Battelier B, Baynham C~F~A, Beaufils Q, Beli{\'{c}} A, Berg{\'e} J, Bernabeu J, Bertoldi A, Bingham R, Bize S, Blas D, Bongs K, Bouyer P, Braitenberg C, Brand C, Braxmaier C, Bresson A, Buchmueller O, Budker D, Bugalho L, Burdin S, Cacciapuoti L, Callegari S, Calmet X, Calonico D, Canuel B, Caramete L~I, Carraz O, Cassettari D, Chakraborty P, Chattopadhyay S, Chauhan U, Chen X, Chen Y~A, Chiofalo M~L, Coleman J, Corgier R, Cotter J~P, Michael~Cruise A, Cui Y, Davies G, De~Roeck A, Demarteau M, Derevianko A, Di~Clemente M, Djordjevic G~S, Donadi S, Dor{\'e} O, Dornan P, Doser M, Drougakis G, Dunningham J, Easo S, Eby J, Elertas G, Ellis J, Evans D, Examilioti P, Fadeev P, Fan{\`i} M, Fassi F, Fattori M, Fedderke M~A, Felea D, Feng C~H, Ferreras J, Flack R, Flambaum V~V, Forsberg R, Fromhold M, Gaaloul N, Garraway B~M, Georgousi M, Geraci A, Gibble K,
  Gibson V, Gill P, Giudice G~F, Goldwin J, Gould O, Grachov O, Graham P~W, Grasso D, Griffin P~F, Guerlin C, G{\"u}ndo{\u{g}}an M, Gupta R~K, Haehnelt M, Han{\i}meli E~T, Hawkins L, Hees A, Henderson V~A, Herr W, Herrmann S, Hird T, Hobson R, Hock V, Hogan J~M, Holst B, Holynski M, Israelsson U, Jegli{\v{c}} P, Jetzer P, Juzeli{\={u}}nas G, Kaltenbaek R, Kamenik J~F, Kehagias A, Kirova T, Kiss-Toth M, Koke S, Kolkowitz S, Kornakov G, Kovachy T, Krutzik M, Kumar M, Kumar P, L{\"a}mmerzahl C, Landsberg G, Le~Poncin-Lafitte C, Leibrandt D~R, L{\'e}v{\`e}que T, Lewicki M, Li R, Lipniacka A, Lisdat C, Liu M, Lopez-Gonzalez J~L, Loriani S, Louko J, Luciano G~G, Lundblad N, Maddox S, Mahmoud M~A, Maleknejad A, March-Russell J, Massonnet D, McCabe C, Meister M, Me{\v{z}}nar{\v{s}}i{\v{c}} T, Micalizio S, Migliaccio F, Millington P, Milosevic M, Mitchell J, Morley G~W, M{\"u}ller J, Murphy E, M{\"u}stecapl{\i}o{\u{g}}lu {\"O}~E, O'Shea V, Oi D~K~L, Olson J, Pal D, Papazoglou D~G, Pasatembou E, Paternostro M,
  Pawlowski K, Pelucchi E, Pereira~dos Santos F, Peters A, Pikovski I, Pilaftsis A, Pinto A, Prevedelli M, Puthiya-Veettil V, Quenby J, Rafelski J, Rasel E~M, Ravensbergen C, Reguzzoni M, Richaud A, Riou I, Rothacher M, Roura A, Ruschhaupt A, Sabulsky D~O, Safronova M, Saltas I~D, Salvi L, Sameed M, Saurabh P, Sch{\"a}ffer S, Schiller S, Schilling M, Schkolnik V, Schlippert D, Schmidt P~O, Schnatz H, Schneider J, Schneider U, Schreck F, Schubert C, Shayeghi A, Sherrill N, Shipsey I, Signorini C, Singh R, Singh Y, Skordis C, Smerzi A, Sopuerta C~F, Sorrentino F, Sphicas P, Stadnik Y~V, Stefanescu P, Tarallo M~G, Tentindo S, Tino G~M, Tinsley J~N, Tornatore V, Treutlein P, Trombettoni A, Tsai Y~D, Tuckey P, Uchida M~A, Valenzuela T, Van Den~Bossche M, Vaskonen V, Verma G, Vetrano F, Vogt C, von Klitzing W, Waller P, Walser R, Wille E, Williams J, Windpassinger P, Wittrock U, Wolf P, Woltmann M, W{\"o}rner L, Xuereb A, Yahia M, Yazgan E, Yu N, Zahzam N, Zambrini~Cruzeiro E, Zhan M, Zou X, Zupan J and
  Zupani{\v{c}} E 2022 {\em EPJ Quantum Technology\/} {\bf 9} 30 ISSN 2196-0763 \urlprefix\url{https://doi.org/10.1140/epjqt/s40507-022-00147-w}

\bibitem{DaRos2023}
Da~Ros E, Kanthak S, Sa\u{g}lamy\"urek E, G\"undo\u{g}an M and Krutzik M 2023 {\em Phys. Rev. Res.\/} {\bf 5}(3) 033003 \urlprefix\url{https://link.aps.org/doi/10.1103/PhysRevResearch.5.033003}

\bibitem{Rohringer2008}
Rohringer W, B{\"{u}}cker R, Manz S, Betz T, Koller C, G{\"{o}}bel M, Perrin A, Schmiedmayer J and Schumm T 2008 {\em Applied Physics Letters\/} {\bf 93} 1--4 ISSN 00036951 (\textit{Preprint} \eprint{0810.4474})

\bibitem{Geisel2013}
Geisel I, Cordes K, Mahnke J, J{\"{o}}llenbeck S, Ostermann J, Arlt J, Ertmer W and Klempt C 2013 {\em Applied Physics Letters\/} {\bf 102} ISSN 00036951 (\textit{Preprint} \eprint{1305.4094})

\bibitem{Tranter2018}
Tranter A~D, Slatyer H~J, Hush M~R, Leung A~C, Everett J~L, Paul K~V, Vernaz-Gris P, Lam P~K, Buchler B~C and Campbell G~T 2018 {\em Nature Communications\/} {\bf 9} ISSN 20411723 (\textit{Preprint} \eprint{1805.00654}) \urlprefix\url{http://dx.doi.org/10.1038/s41467-018-06847-1}

\bibitem{Wu2020}
Wu Y, Meng Z, Wen K, Mi C, Zhang J and Zhai H 2020 {\em Chinese Physics Letters\/} {\bf 37} ISSN 17413540 (\textit{Preprint} \eprint{2003.11804})

\bibitem{reinschmidt2023}
Reinschmidt M, Fort{\'{a}}gh J, G{\"{u}}nther A and Volchkov V 2023 Reinforcement learning in ultracold atom experiments (\textit{Preprint} \eprint{2306.16764})

\bibitem{Wigley2016}
Wigley P~B, Everitt P~J, {Van Den Hengel} A, Bastian J~W, Sooriyabandara M~A, Mcdonald G~D, Hardman K~S, Quinlivan C~D, Manju P, Kuhn C~C, Petersen I~R, Luiten A~N, Hope J~J, Robins N~P and Hush M~R 2016 {\em Scientific Reports\/} {\bf 6} 1--6 ISSN 20452322 (\textit{Preprint} \eprint{1507.04964}) \urlprefix\url{http://dx.doi.org/10.1038/srep25890}

\bibitem{Nakamura2019}
Nakamura I, Kanemura A, Nakaso T, Yamamoto R and Fukuhara T 2019 {\em Optics Express\/} {\bf 27} 20435 ISSN 1094-4087

\bibitem{Barker2020}
Barker A~J, Style H, Luksch K, Sunami S, Garrick D, Hill F, Foot C~J and Bentine E 2020 {\em Machine Learning: Science and Technology\/} {\bf 1} 015007 ISSN 2632-2153 (\textit{Preprint} \eprint{1908.08495}) \urlprefix\url{https://iopscience.iop.org/article/10.1088/2632-2153/ab6432}

\bibitem{Davletov2020}
Davletov E~T, Tsyganok V~V, Khlebnikov V~A, Pershin D~A, Shaykin D~V and Akimov A~V 2020 {\em Physical Review A\/} {\bf 102} 11302 ISSN 24699934 (\textit{Preprint} \eprint{2003.00346}) \urlprefix\url{https://doi.org/10.1103/PhysRevA.102.011302}

\bibitem{Ma2023}
Ma J, Fang R, Han C, Jiang X, Qiu Y, Ma Z, Wu J, Zhan C, Li M, Lu B and Lee C 2023  (\textit{Preprint} \eprint{2303.05358}) \urlprefix\url{http://arxiv.org/abs/2303.05358}

\bibitem{poli2007}
Poli R, Kennedy J and Blackwell T 2007 {\em Swarm intelligence\/} {\bf 1} 33--57

\bibitem{kaveh2014}
Kaveh A and Zolghadr A 2014 {\em Computers \& Structures\/} {\bf 130} 10--21

\bibitem{CMAES}
{CMA-ES, Covariance Matrix Adaptation Evolution Strategy for non-linear numerical optimization in Python} \url{https://pypi.org/project/cma/}

\bibitem{nelder1965}
Nelder J~A and Mead R 1965 {\em The computer journal\/} {\bf 7} 308--313

\bibitem{AX}
{Bayesian Optimization} \url{https://ax.dev/}

\bibitem{FMFN}
Nogueira F 2014-- {Bayesian Optimization}: Open source constrained global optimization tool for {Python} \url{https://github.com/bayesian-optimization/BayesianOptimization} \urlprefix\url{https://github.com/fmfn/BayesianOptimization}

\bibitem{letham2019}
Letham B, Karrer B, Ottoni G and Bakshy E 2019 {\em Bayesian Analysis\/} {\bf 14} 495 -- 519 \urlprefix\url{https://doi.org/10.1214/18-BA1110}

\bibitem{optimal}
{OptimalQT} \urlprefix\url{https://jcmwave.com/company/projects/item/1052-optimal-qt}

\bibitem{zhang2015}
Zhang Y, Wang S, Ji G {\em et~al.\/} 2015 {\em Mathematical problems in engineering\/} {\bf 2015}

\bibitem{das2010}
Das S and Suganthan P~N 2010 {\em IEEE transactions on evolutionary computation\/} {\bf 15} 4--31

\bibitem{hansen2006}
Hansen N 2006 {\em Towards a new evolutionary computation: Advances in the estimation of distribution algorithms\/}  75--102

\bibitem{shahriari2015}
Shahriari B, Swersky K, Wang Z, Adams R~P and De~Freitas N 2015 {\em Proceedings of the IEEE\/} {\bf 104} 148--175

\bibitem{rasmussen2006}
Rasmussen C~E, Williams C~K {\em et~al.\/} 2006 {\em Gaussian processes for machine learning\/} vol~1 (Springer)

\bibitem{bull2011}
Bull A~D 2011 {\em Journal of Machine Learning Research\/} {\bf 12}

\bibitem{picheny2013}
Picheny V, Wagner T and Ginsbourger D 2013 {\em Structural and multidisciplinary optimization\/} {\bf 48} 607--626

\bibitem{schneider2019}
Schneider P~I, Garcia~Santiago X, Soltwisch V, Hammerschmidt M, Burger S and Rockstuhl C 2019 {\em ACS Photonics\/} {\bf 6} 2726--2733

\bibitem{ColdQuanta2023}
ColdQuanta {RuBECi} \urlprefix\url{https://www.shopcoldquanta.com/rubeci}

\bibitem{Wicht2017}
Wicht A, Bawamia A, Kr{\"{u}}ger M, K{\"{u}}rbis C, Schiemangk M, Smol R, Peters A and Tr{\"{a}}nkle G 2017  {\bf 10085} 100850F ISSN 1996756X \urlprefix\url{http://proceedings.spiedigitallibrary.org/proceeding.aspx?doi=10.1117/12.2253655}

\bibitem{Pical}
{Pical - Picture analysis for cold atoms } \urlprefix\url{https://git.physik.hu-berlin.de/pical/pical-picture-analysis-for-cold-atoms}

\bibitem{M-labs2023}
M-labs Sinara hardware \urlprefix\url{https://m-labs.hk/experiment-control/sinara-core/}

\end{thebibliography}

\end{document}